\documentstyle[preprint,eqsecnum,aps,epsfig]{revtex}

\begin{document}
\draft
\preprint{ }
\title{ Measurement of air shower cores to  study  the cosmic ray composition 
in the knee energy region
}
\author{
 M.~Amenomori$^{1}$, S.~Ayabe$^{2}$,   Caidong$^{3}$,  Danzengluobu$^{3}$,
 L.K.~Ding$^{4}$,   Z.Y.~Feng$^{5}$,   Y.~Fu$^{6}$,    H.W.~Guo$^{3}$, 
 M.~He$^{6}$,  K.~Hibino$^{7}$,   N.~Hotta$^{8}$,    Q.~Huang$^{5}$,
 A.X.~Huo$^{4}$,  K.~Izu$^{9}$,  H.Y.~Jia$^{5}$,    F.~Kajino$^{10}$, 
 K.~Kasahara$^{11}$,   Y.~Katayose$^{12}$,  Labaciren$^{3}$, J.Y.~Li$^{6}$,
 H.~Lu$^{4}$,   S.L.~Lu$^{4}$,  G.X.~Luo$^{4}$, X.R.~Meng$^{3}$, 
 K.~Mizutani$^{2}$,  J.~Mu$^{13}$,  H.~Nanjo$^{1}$, M.~Nishizawa$^{14}$,
 M.~Ohnishi$^{9}$, I.~Ohta$^{8}$,  T.~Ouchi$^{7}$,  Z.R.~Peng$^{4}$,       
  J.R.~Ren$^{4}$,  T.~Saito$^{15}$,  M.~Sakata$^{10}$, T.~Sasaki$^{10}$, 
  Z.Z.~Shi$^{4}$, M.~Shibata$^{12}$,   A.~Shiomi$^{9}$,  T.~Shirai$^{7}$,
  H.~Sugimoto$^{16}$,  K.~Taira$^{16}$,  Y.H.~Tan$^{4}$,
  N.~Tateyama$^{7}$,    S.~Torii$^{7}$,  T.~Utsugi$^{2}$,  C.R.~Wang$^{6}$,
  H.~Wang$^{4}$,   X.W.~Xu$^{4,9}$,       Y.~Yamamoto$^{10}$,  G.C.~Yu$^{3}$, 
  A.F.~Yuan$^{3}$, T.~Yuda$^{9,17}$,    C.S.~Zhang$^{4}$,    H.M.~Zhang$^{4}$, 
  J.L.~Zhang$^{4}$,   N.J.~Zhang$^{6}$,  X.Y.~Zhang$^{6}$,
   Zhaxiciren$^{3}$, and   Zhaxisangzhu$^{3}$ \\
(The Tibet AS${\bf \gamma}$ Collaboration)
} 
\address{
$^{1}$Department of Physics, Hirosaki University, Hirosaki 036-8561, Japan\\
$^{2}$Department of Physics, Saitama University, Urawa 338-8570, Japan\\
$^{3}$Department of Mathematics and Physics, Tibet University, Lhasa 850000, China\\
$^{4}$Laboratory of Cosmic Ray and High Energy Astrophysics,\\
      Institute of High Energy 
          Physics, Academia Sinica, Beijing 100039, China\\
$^{5}$ Department of Physics, South West Jiaotong University, Chengdu 610031, China\\
$^{6}$ Department of Physics, Shangdong University, Jinan 250100, China\\
$^{7}$ Faculty of Engineering, Kanagawa University, Yokohama 221-8686, Japan\\
$^{8}$ Faculty of Education, Utsunomiya University, Utsunomiya 321-8505, Japan\\
$^{9}$ Institute for Cosmic Ray Research, University of Tokyo, Kashiwa 277-8582, Japan\\
$^{10}$ Department of Physics, Konan University, Kobe 658-8501, Japan\\
$^{11}$ Faculty of Systems Engineering, Shibaura Institute of Technology, 
Omiya 330-8570, Japan\\
$^{12}$ Faculty of Engineering, Yokohama National University, Yokohama 240-0067, Japan\\
$^{13}$ Department of Physics, Yunnan University, Kunming 650091, China\\
$^{14}$ National Institute of Informations, Tokyo 101-8430, Japan\\
$^{15}$ Tokyo Metropolitan College of Aeronautical Engineering, Tokyo 116-0003, Japan\\
$^{16}$ Shonan Institute of Technology, Fujisawa 251-8511, Japan\\
$^{17}$ Solar-Terrestrial Environment Laboratory, Nagoya University, Nagoya 464-8601, Japan
 }
\date{\today}
\maketitle
\begin{abstract} 

Since 1996, a hybrid experiment consisting of an emulsion chamber
and a burst detector array and the Tibet-II air shower array
has been operated at Yangbajing (4300~m above sea level) in Tibet.
This experiment can detect air-shower cores, called burst
events, accompanied by air showers in excess of about 100~TeV. 
Using the burst event data observed by this experiment,
we discuss the primary cosmic ray composition around the knee
in comparison with the Monte Carlo simulations.
In this paper, we show that all the features of  burst events
are wholly compatible with the heavy enriched composition in the
knee energy region.
\end{abstract}

\pacs{PACS numbers : 95.85Ry, 98.70Sa, 96.40Pq, 96.40De}

\section{INTRODUCTION}

It is commonly understood that the knee of the primary cosmic ray
spectrum has its origin in the acceleration and propagation
of high energy cosmic rays in the Galaxy. The model of the shock 
acceleration by supernova
blast waves leads to the formation of a power-law spectrum of 
particle energies with a differential
index of about -2 at sources \cite{1}. The plausible propagation 
models of their confinement by
galactic magnetic fields and of their eventual escape from our Galaxy
can explain well a steeper power-law spectrum than that
at the source region \cite{2}, suggesting a rigidity-dependent 
bending for different cosmic ray composition.
Within the framework of this picture the average mass of primary cosmic
rays before the knee should increase with increasing primary energy. 
In other words, the knee composition 
should be heavily dominant as the proton spectrum may bend at an energy of  
about 100~TeV corresponding to a maximum energy gained by shock 
acceleration at supernova remnants.

While there is no consensus on the origin of cosmic rays with energies beyond the knee,
observations of cosmic rays in such a high energy region may naturally stand in need of
another acceleration mechanisms \cite{3,4} or new cosmic 
ray sources \cite{5,6}. Among those, one of the most promising  models
may be  that the cosmic rays come
from extra-galactic sources such as active galactic nuclei \cite{6},
although the evidence is far from convincing. However, such an 
extra-galactic source model should  predict proton enriched primary composition 
around and beyond the knee.
   
Thus, measurements of the primary cosmic rays around the knee
are very important and its composition is a fundamental input for understanding the 
particle acceleration mechanism that pushes cosmic rays to very high energies. 
However, because of extremely low and steeply decreasing flux at high energies, 
 direct measurements of primary cosmic rays on board balloons are 
still limited in the energy region to below a few hundred TeV \cite{7,8}.
To date, the knee composition of primary cosmic rays has been studied 
by observing  air showers with a large aperture ground-based apparatus. 
For air shower observations, a surface detector array is commonly 
set up to measure the lateral distributions and the arrival times
of shower particles (mostly electrons and photons), which enable us to 
locate the core position, determine the arrival direction, and 
deduce the shower size and shower age \cite{9}.  These are important basic 
parameters to describe an air shower,  although they are not very sensitive
to the primary composition. In order to study the composition around the
knee,  measurements of muon
content in each air shower \cite{10} or muons in deep underground \cite{11,12}, 
measurements of lateral distribution of air shower Cherenkov lights \cite{13}, or 
maximum depth of shower development using air Cherenkov telescopes \cite{14},  
and  multiparameter measurements of air showers \cite{15}
have been carried out and devoted to making the conclusion about the knee 
composition.  In spite of great efforts so far, however, there is a divergence of 
conclusions on the composition  from experiment to experiment and the knee composition 
is still the question at issue. Another approach may be required to get more 
direct information about the composition.

Within the ground-based air shower experiments those set up at higher altitudes
are preferable for the physics studies at the knee region. The reasons include ; 
first,  maximum development of showers with the knee
energies is  closer to the observation level, so that the shower is less
fluctuated and the energy
determination is more precise and less dependent upon the unknown
composition \cite{9} ; second, the energy flow 
in the core region of air showers is less attenuated and easier to observe
with conventional calorimeters. High energy air-shower cores are sensitive
to the composition of the primary cosmic rays around the knee.
Air-shower cores contain a large part of the primary energy in the early stage of
shower development, but with increasing atmospheric depth 
they are rapidly diffused with dissipation of  their energies.  Hence,
the measurement of air-shower cores should be implemented at high
altitude with a detector having a reasonably large area. In general a thick
shower detector is required when one wants to record all (or most) hadronic 
components in the core, but this needs enormous heavy materials to
absorb them in the detector. At high altitude, however, this can be achieved by
observing the electromagnetic components in the core with a thin detector,
since these  are mostly the cascade products induced  by high energy $\pi^0$-decay
$\gamma$ rays  which are produced in the air shower cores. Therefore,
the electromagnetic components in the core well reflect the major behavior
of the whole hadronic components,  keeping the sensitivity to the primary composition.

In this paper we report our study of the primary composition
in the knee region using data of the Tibet burst detector and 
the air-shower array II.   The experiment, including
the apparatus and its performance,  data set and background analysis, 
is introduced in Sec. II.
 The air-shower simulation and detector response calculation
are described in Sec. III. The results and discussions
 are described in Sec. IV.
A brief summary is given in Sec, V.

\section{EXPERIMENT}

\subsection {Apparatus}

We started a hybrid experiment of the emulsion chamber, the burst detector 
and the air shower array  (Tibet-II) at Yangbajing (4300 m above sea 
level), Tibet in 1996 \cite{16}.  The Tibet-II array
consists of 221 scintillation counters of 0.5 m$^2$ each which are placed on
a 15 m square grid and has been
operated since 1995. Any fourfold coincidence in the detectors 
is used as the trigger 
condition for air-shower events. Under this condition the trigger rate
is about 200 Hz with a dead time of about 12\%  for data taking. The energy 
threshold is estimated 
to be about 7 TeV for proton induced showers. The precision of the
shower direction determination is about $1^\circ$, which has been confirmed
by observing the Moon's shadow \cite{17}. The main aim of Tibet-II is
to search for  $\gamma$-ray point sources at energies around 10 TeV. But it 
can also be used for
the measurement of the spectrum of the total cosmic ray particles \cite{9}, and for
the study of topics in the knee region by providing information
of the shower size, direction, core position, and arrival time
of each air-shower event to the core detectors \cite{16}.

\begin{figure}
\begin{center}
 \vspace{2cm}
  \epsfig{file=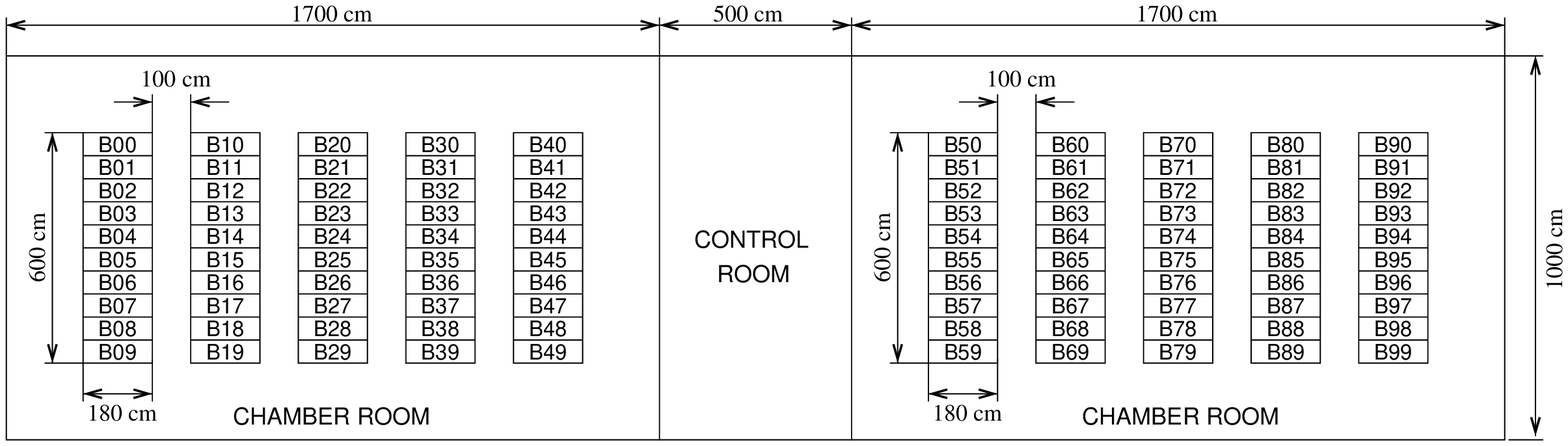,height=4.5cm} \par
\end{center}
\caption{ Arrangement of the burst detectors in two rooms. The area of
each burst detector is 50 cm $\times$ 160 cm and four emulsion chambers
are set up on each burst detector.}
\label{Fig1}
\end{figure}

\begin{figure}
\begin{center}
 \vspace{2cm}
  \epsfig{file=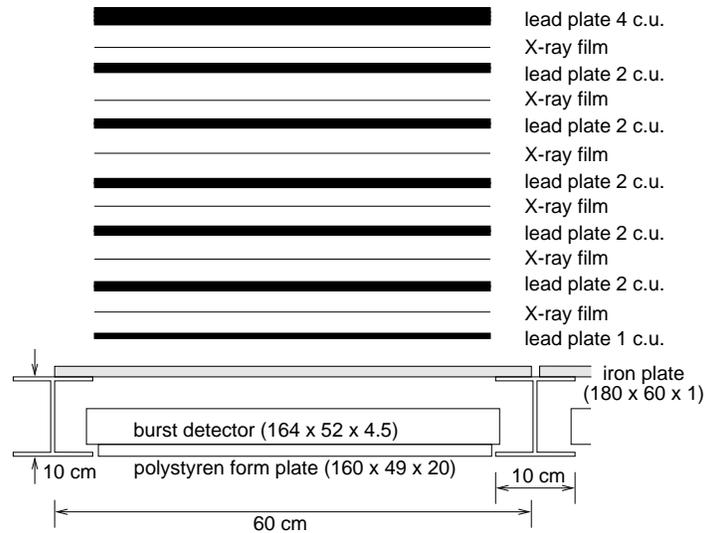,height=9cm} \par
\end{center}

\caption{ Schematic side view of each unit of emulsion chamber and 1/4
of 1 unit burst
detector. High sensitive x-ray films are inserted at every 2 c.u. in
emulsion. Total thickness of lead plates is 14 c.u.}
\label{Fig2}
\end{figure}

The emulsion chambers and the burst detectors are used to detect high-energy air-shower cores 
accompanied by air showers induced by primary cosmic rays with energies above 10$^{14}$ eV. 
They are set up separately in two rooms as shown in Fig. \ref{Fig1}
and placed near the center of the Tibet-II array. 
A basic structure of each  emulsion chamber used here is a multilayered
sandwich of lead plates and photosensitive
x-ray films. Photosensitive layers are put every 2 cascade unit (c.u.) (1 c.u. = 0.5 cm) of lead in the
chamber as shown in Figure \ref{Fig2}.
 There are 400 units of emulsion chamber, each with an area of
40 cm $\times$  50 cm with the total thickness of 15 c.u., giving the total sensitive area of 80 m$^2$, and 100
units of burst detectors  each with an effective
area of 160 cm $\times$ 50 cm. Four units of the emulsion chamber are put above
one unit of the burst detector. A 1 cm iron plate is put between emulsion
chambers and burst detectors.

Each burst detector consists of a plastic scintillator with the size of
160 cm $\times$ 50 cm and thickness of 2 cm, and four photodiodes (PDs) are
 attached at four
corners of each scintillator to read light signals generated by shower
particles produced in the lead and iron absorber above the
detector. From the analogue-to-digital converter (ADC) values of four PDs the total number
(i.e., burst size $N_b$) and the position of the number-weighted center of all
shower particles that hit a burst detector can be estimated. The 
response of the burst detector is calibrated using electron beams from 
an accelerator. The performance of the burst detector and the calibration using
the electron beams are briefly summarized in the Appendix. It is confirmed that
the measurable shower size by each burst detector ranges from $10^4$ to
$3\times10^6$, roughly corresponding to showers with energies ranging 
from several times 100 GeV to about 300 TeV. A burst event is triggered
when any two-fold coincidence of signals from four PDs of a burst
detector appears. The coincidence of a burst event and an air-shower event
 is made by their arrival times, and the coincidence of a burst event and
a family event observed in the emulsion chamber is made by their positions
and directions. (A burst event and its accompanying air shower have
the same direction.)

In this analysis we  use only the data obtained from all burst detectors and
the Tibet-II array, while the emulsion chamber data will be
reported elsewhere in the very near future.  Using the 
burst detector array  shown in Fig. \ref{Fig1}, the electromagnetic
 components in the air-shower cores can be measured in the area within 
a radius of several meters.

\subsection{Data set}

The data set of the burst detector used for this work was taken 
from 16 October 1996 to 1 June 1999.
First we scanned  the target maps of all events with the naked eye.  All those events
showing a systematic noise configuration were ruled out during the
first scanning. Then we removed  the events that were not coincident with
the air-shower events recorded by the Tibet-II array. Finally we imposed the
 following conditions on the events :   
(1) when the number of fired detectors is only one, its size $N_b$ should be larger than
$5\times10^4$ ; and (2) when more than one detector is fired,  the largest size 
(hereafter for each event we call the detector that  observed the highest burst size
the TOP detector) should be larger than $5\times10^4$ and also the size of any other 
one should be larger than $3\times10^4$ (minimum size). 

\begin{figure}

\begin{center}
 \vspace{2cm}
  \epsfig{file=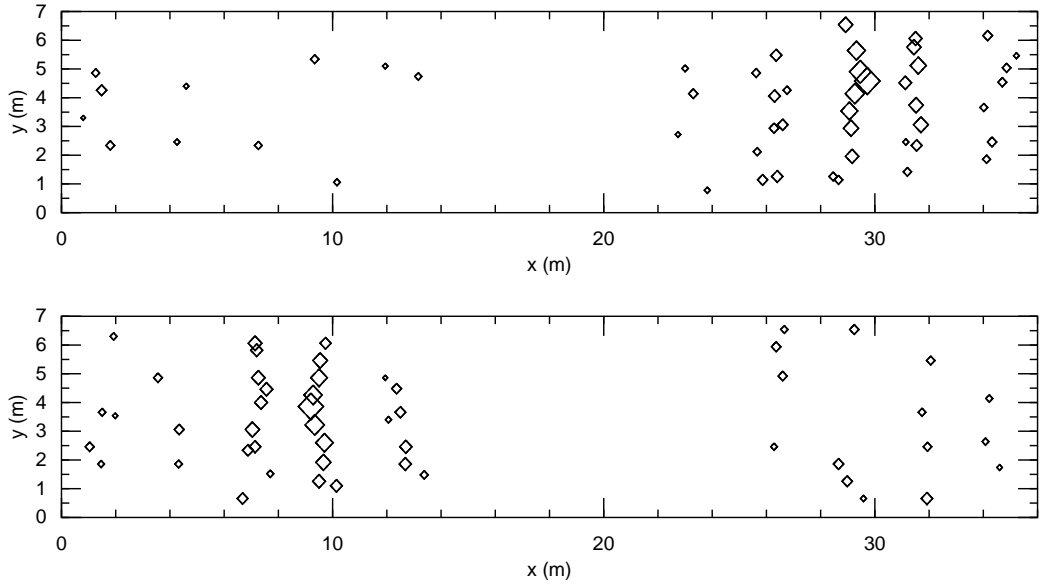,height=8cm} \par
\end{center}

\caption{Two examples of air-shower core events observed in the burst detectors.
Rhombi denote the size of events observed in each burst detector and its
geometrical size is logarithmically proportional to the burst size.}
\label{Fig3}
\end{figure}

 From this procedure, we selected  9278 events in total.  
 Two examples of the burst detector events are shown in Fig. \ref{Fig3},  where the
scale of the marks is logarithmically proportional to the burst
size. A remarkable core structure in the event pattern may be recognized.

\subsection{Background analysis}

\begin{figure}

\begin{center}
  \epsfig{file=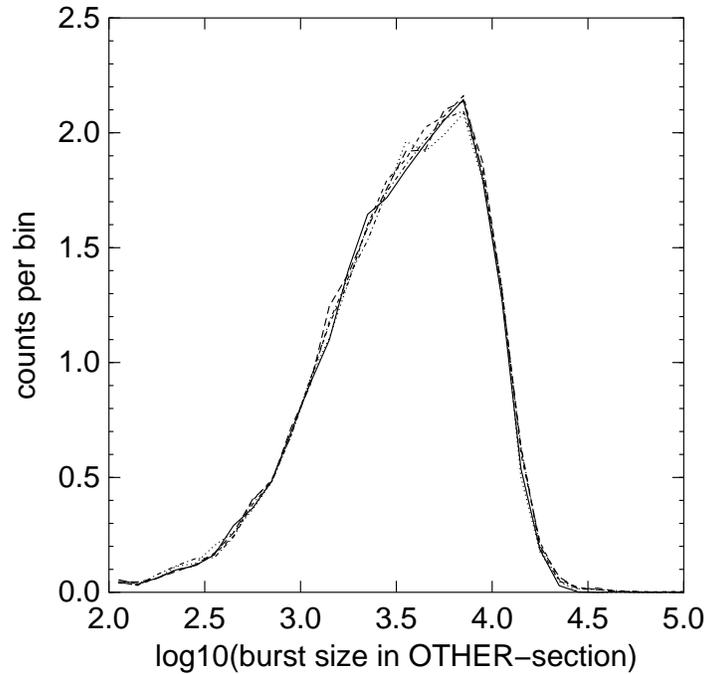,height=9cm} \par
\end{center}
\caption{ Burst size distribution of any one detector in the OTHER section. The five curves correspond
to the different positions of the TOP detector being in the first, second, third, fourth
and fifth column of the TOP section, respectively.}
\label{Fig4}
\end{figure}

 We  carefully examined  whether some background exists in the data of each event
since the whole burst detector is separated into two sections
and there is a distance of 9 m between them.  In the following, we call
 the section containing the
TOP detector  as the ^^ ^^ TOP section'', and the other as the
^^ ^^ OTHER section''. We first examined whether those bursts located
far from the TOP detector still contain signals or not. For this, we divided
all events into five groups according to the TOP detector in which each event was
in the first, second, third, fourth and fifth column of the TOP section, and then the 
size distribution of any one burst
detector in the OTHER section was obtained, respectively, for each group.
It is expected that if the bursts recorded in the OTHER section 
contain signals, their burst size distribution  should be
different with different event groups because they  have 
different core distances. However 
these five distributions, which are taken from the first year's data set,
 are  almost the same as seen in Fig. \ref{Fig4},
showing to be independent of the  core distance.
This suggests that when the distance to the TOP detector (i.e., to the
air-shower core), is larger than 10 m almost all recorded bursts are
formed by noises ranging from $10^2$ to $3 \times 10^4$ under our experimental conditions.
The same burst size distributions in the OTHER section are seen from the data sets of 
the second and third year's observation, showing that a stable noise existed 
during the whole period of the operation. 
Therefore, the distribution showing in Fig. 4 is recognized as the 
background coming from our experimental conditions. These noises may
 be mostly induced by an incomplete ground connection of the detectors to the earth.
Actually, when the thunder is rumbling, an increase of the noises is observed.
Hence, it is reasonable 
to assume that some background may still exist in the data of the TOP section
although the minimum shower size for the burst detector is set to be $3 \times 10^4$ in the present analysis. The treatment of the background is discussed 
later when the data are compared with the simulation.

\section{MONTE CARLO SIMULATION}

\subsection{Air-shower simulation}

Air showers induced by different primary particles were generated by using
 the Monte Carlo codes CORSIKA-QGSJET \cite{18} and COSMOS  \cite{19} 
which  have 
been used for many cosmic ray experiments and shown to be able to explain  many
quantities at the energy region below and around the knee fairly well. 
The all-particle spectrum measured by the Tibet AS$\gamma$
Collaboration  \cite{9} is used as the input of simulations. The minimum
energy of primary particles to be sampled  is taken to be 500 TeV 
and the zenith angle at the top of the atmosphere is uniformly sampled 
between  $0^\circ$ and $45^\circ$. Since the chemical composition of
primary particles is unknown around the knee region, four different
composition models are examined to compare with the experiment.
While two of them, being pure protons and pure irons,  are extreme assumptions, 
these  provide  some  boundary of predictions used  as the first check
of the interaction models adopted in the Monte Carlo simulation.
When some data  happen to be outside the boundary,  this may
raise some points in the interaction models 
that must be ruled out. 
 Two others are
the heavy dominant (HD) and  proton dominant (PD) models  \cite{20}, 
which are shown in  Fig. \ref{Fig5}.
In both models the chemical composition is divided into seven groups : 
proton, helium, 
light ($L, Z=8$), medium ($M, Z=14$), heavy ($H, Z=25$), very heavy ($VH, Z=35$) 
and iron. The HD model may be related to a supernova acceleration and 
rigidity dependent propagation model in that the proton component is 
assumed to bend at energy of about 100 TeV.
The fraction of the iron component increases with increasing primary energy, resulting
in the primary becoming heavy dominant at the knee. 
 The PD model assumes a proton dominant chemical composition
 over the whole knee energy region, while all components 
bend at energies of 2000 TeV.  
In both HD and PD models the summation of all individual spectra of
different nuclear species is fitted to  the observed total particle spectrum 
as shown in Fig. \ref{Fig5}.

In this simulation all shower particles are followed 
by a full Monte Carlo method until their energies become 1 GeV. 
Although the smallest burst size to be observed
per detector is taken to be $3 \times 10^4$ (see Sec. II B), corresponding
to a few to 10 TeV for a single $\gamma$-ray or a single electron
incident on the surface of the burst detector, 
we need to follow shower particles untill 1 GeV since a large number
of low energy particles near the
shower core hitting the detector make a contribution to the 
observed burst size that is not negligible.\footnote{Setting the
lowest energy at 1 GeV does not mean the observational threshold of
our burst detector is as low as 1 GeV. The major contribution to the
burst events as we analyzed in this work comes from particles with
energy higher than 100 GeV.}
However, we determined  that a contribution from shower particles lower  
than 1 GeV is insignificant, i.e., smaller than $1\%$.

\subsection{Simulation for burst detector}

 Shower developments in the burst detector were calculated based on the data 
obtained using a Monte Carlo code, EPICS \cite{21}.
 When an air-shower event reaches the observation level it
is dropped within the area of 14 m$\times$21 m (294 m$^2$) for that  
one section of the whole burst detector is assumed to be located
in its central part. For each air-shower particle with energy higher than 1 GeV
(electrons, positrons and $\gamma$ rays. Further interactions of pions and muons in the
detector are neglected.) 
its cascade development in the detector is calculated analytically
and then the number of cascade particles just below  the lead and iron
plates  is obtained. The analytical formulae used here were made by
modifying the well known cascade functions \cite{22} and the parameters involved in the
formulae  were adjusted by using the data from a set of Monte Carlo 
events generated by EPICS. 
Fluctuations of the number of cascade particles are adequately
 taken into account.
Air-shower cores enter at various positions of the burst detector and then charged 
particles
passing through the scintillator emit photons. 
Photons  are assumed to attenuate in the scintillator as $ r^{-1.2}$,
 where $r$ is the distance between the shower hit position and one of the
four PDs (see Appendix).
In this detector simulation, these are taken into account to get an ADC
signal from each photodiode. A  size $N_b$ and its hit position of burst event
in each detector are  then estimated with the
same procedure as the experiment. Because of the saturation of ADC 
outputs, the detectable size per burst detector is limited to be
smaller than about $3 \times 10^6$.

\subsection{Background treatment}

As shown in the previous section,  there exist some noises in the data, 
 so their effects must be carefully 
taken into account.
 We also found that for the experimental data it is difficult to subtract 
the background in the TOP section in a correct  
way. Then an opposite approach is adopted here. That is, in order to
compensate the effects from the background we added the experimental
background to the simulation samples as follows: for any one of the
simulated events,
a background event is randomly taken from the experimental background
data set that contains about $10^4$ events, and then this background event
is added to the simulated event at their corresponding
positions of the noise detectors. 

\section{RESULTS AND DISCUSSIONS}

\subsection{Event selection}

For a further analysis, we selected the events from both the experimental data
 set
and the Monte Carlo data set by imposing the following criteria (for convenience
hereafter we call it ^^ ^^ criteria-A'') : (1) Zenith angle $\leq 45^\circ$ ; (2)
 burst size of the TOP detector, $N_b^{top} \geq  5 \times 10^4$ ;  (3)
 burst size of any non-TOP detector, $N_b^{non-top} \geq$ $3 \times 10^4$ ; and (4)
  number of fired detectors, $N_{bd} \geq 4$.

By this selection 1046 events are obtained from the experimental data set. 
The time interval between two neighboring events is examined and an
exponential-type distribution is confirmed, indicating a good randomness
of this data sample. The effective running time of this sample is estimated
to be 7.54 $\times10^7$ s. Also, as we  adopt only the burst events which
are coincident with the air-shower events recorded by the Tibet-II array,
a dead time of 12 \% for data taking of the array must be taken into
account. Hence,  when we discuss the flux of the burst events, we should use
1172 ( = 1046 $\times$ 1.12) as the number of burst events satisfying criteria-A.

For the simulations, as mentioned above, we generated air-shower events
using different primary assumptions for which the primary particles were
sampled from the same all-particle spectrum starting from 500 TeV. 14810,
9143, 8463 and 10591 events satisfying  criteria-A were obtained
by the composition assumptions of pure protons, PD, HD, and pure irons,
respectively, which are about 8 - 14 times larger than the experimental
data set. Figure \ref{Fig6} shows the energy distributions of the
 primary particles responsible for generating 
the samples of selected burst events for these four assumptions. 
It is 
seen that the mode energy of primary protons capable of generating
the selected burst events is about 2500 TeV, while about 5000 TeV
for irons. Therefore, the burst event samples satisfying criteria-A
can easily  manifest the behavior of primary particles in the knee region.

\subsection{Flux of the burst events}

We first discuss the primary composition from the point of the 
intensity of the burst events
satisfying criteria-A.
 
Using the all-particle spectrum obtained by the Tibet air-shower experiment,
the number of primary particles with energies above 10$^{16}$ eV and with zenith angles less than 45$^\circ$, which fall within the effective burst-detector
 area of
 2 $\times$ 294 m$^2$ (corresponding to the effective area of 
two separated rooms)  during the running time of
$7.54 \times 10^7$ s, is calculated to be 1105. From the simulations,
on the other hand, the efficiencies of the primary particles with energy in excess of
10$^{16}$ eV to generate the burst events satisfying  criteria-A 
 are calculated to be 0.33, 0.32, 0.31, and 0.30 for the primary models
of pure protons, PD, HD, and pure irons, respectively. 
Thus,  among the 1105 incidences with energies higher than $10^{16}$ eV,
 365, 354, 343, and 332 events satisfying  criteria-A should be observed
for four primary composition models, respectively. 

The simulation study also tells us that when the composition models of
 pure
protons, PD, HD, and pure irons are assumed, the fractions of the selected burst events induced by the primaries with energies above $10^{16}$ eV to the total selected burst events are calculated to be 0.11, 0.16, 0.25,
 and 0.39, respectively. Accordingly the total selected
events expected to be observed are 
 3438, 2245, 1391 and 845 events, respectively.
These values should be compared with 1,172 events which are truly 
observed with 
the experiment. The ratios of the Monte Carlo expectation to the observation 
are 2.9, 1.9, 1.2 and 0.7 for the four primary models, respectively. 
One can see that the HD model is consistent with the experiment, that is, 
the primary composition around the knee is required to be heavily enriched.

In the following, all the simulation results are normalized
to the experimental one to discuss the behavior of the burst events.
Therefore,  all the distributions 
are given in the ordinate with ^^ ^^ number of events'' that may just correspond 
to the amount of our experimental exposure which is directly related to the
 absolute intensity. 
The error bars of the data in the following figures are statistical ones.

Shown in Figs. \ref{Fig7} and \ref{Fig8} are the total burst size $\sum N_b$ 
distribution, where the summation is done  over all fired detectors satisfying criteria-A
for each event,  and its accompanying air-shower size $N_e$ distribution, respectively. 
A steep slope of the burst size spectrum at their large size region is attributed to the 
saturation effect of ADC outputs from the PDs which are not included in the 
simulation.
 Disregarding a well-ground discrepancy  at their large size region, 
 it is well seen that  the 
intensity of the total burst size spectrum, i.e. the measured total 
electromagnetic component of the air-shower cores  
in the knee region, is compatible with 
 the prediction of HD,   while it is a factor of 2 lower than that 
of PD. A similar situation is also seen in the air-shower size spectrum.

Figure \ref{Fig9} shows the the ratio of $\sum N_b$ to $N_e$. It is seen 
that this 
ratio is sensitive to the primary composition at the region of the ratio $> 1.5$  
and the data are consistent with the HD model.  According to the Monte
Carlo simulation, there is a tendency for
the events with larger $\sum N_b$ and smaller $N_e$  to be mostly
induced by protons and helium nuclei. Incidentally, it may be said that although 
the data are seemingly consistent  with the HD 
model, the saturation effect of the burst size determination (this is not
taken into the simulation) and the to-be-improved air-shower size fit 
seen in Fig. \ref{Fig7} and Fig. \ref{Fig8} may cause some deviation 
from the HD model.
We confirmed, however, that even if such corrections are made the result 
does not change much.

\subsection{Lateral spreads of the burst events}

Figure \ref{Fig10} shows the distribution of $\Sigma r$ of each burst, where $r$ is the 
 the distance (in meters) between  the 
TOP detector's position and the other detector's position, and
the summation is over the whole fired detectors in each
event. A double peak seen in the distribution is due to an oblong
  arrangement of the burst detector units
 with a regular spacing as shown in Fig. \ref{Fig1}.
It is seen that the proton-induced events make more sharp distribution than heavier
nuclei. The experimental
data are also consistent with the HD model.

To examine the sensitivity of the lateral spread of air-shower cores to
the composition, we made the distributions of the value 
$log<N_b>/<r>$ for the experimental and simulated data as shown in Fig.
\ref{Fig11}, where $<N_b>$,
the averaged burst size for one event, is divided by $<r>$, the averaged 
lateral distance between the fired detector and the TOP detector for the 
same event, and the distribution is made over all selected burst events. 
The distributions of the lateral gradient  of burst size defined as 
$[log(N_b^{top})-log(N_b)] / r$ for each burst of all events are    
also presented in 
Figure \ref{Fig12}. 
Both distributions seem to be more sensitive than others to the composition. 
In both cases
the HD composition model can  explain the experiment well.

The number of fired burst detectors $N_{bd}$ also depends on the
 lateral spread of their air-shower core. In  criteria-A
at least four detectors are required to be fired and the average number ranges
from 5.3 to 6.6. The $N_{bd}$ distribution
is shown  in Fig. \ref{Fig13}  and is also  consistent  
with that of the HD model as discussed above.

\subsection{Systematic uncertainties}

We briefly discuss the systematic errors of our results.
The largest one is from the uncertainty of the total primary
flux. We used the all particle spectrum measured by the Tibet
air-shower experiment, and the systematic errors of this result
are estimated to be 20\%-30\% for absolute intensity.
The other may come from the Monte Carlo code we used.
In this study, we used the CORSIKA-QGSJET code.
While  CORSIKA-QGSJET uses a quark-gluon string phenomenology to describe
the hadron-nucleus and nucleus-nucleus interactions, 
COSMOS is another code using a different physics picture ( a Lund string model
 in the lower energy region, but a quasi-scaling assumption  in the higher energy 
region) for the hadronic
and nuclear interactions. It is confirmed that an air-shower simulation using COSMOS+HD
and a detector response calculation using EPICS give almost completely
the same results as CORSIKA+HD. We compare the distributions 
of the total burst size and the total lateral distance, those obtained by using both codes, 
 in Figs. \ref{Fig14} and \ref{Fig15}, respectively.
It is seen that two simulation codes give almost the same results for 
both distributions.
Hence, the results discussed above do not depend on the Monte Carlo code we used.

\section{SUMMARY}

\begin{figure}

\begin{center}
  \vspace{-2cm}  
  \epsfig{file=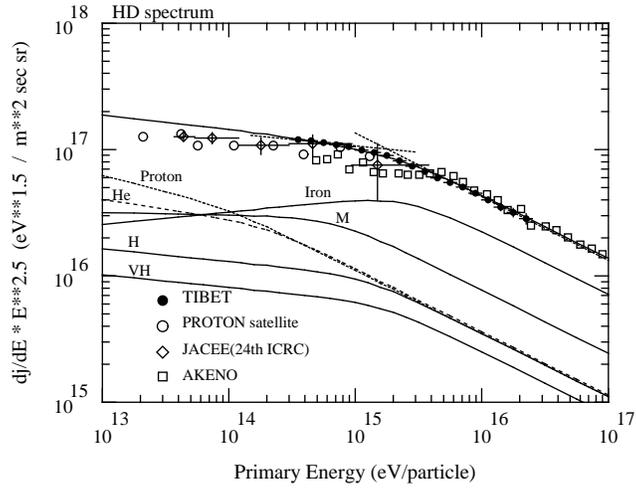,height=11cm} \par
  \vspace{-2cm} 
  \epsfig{file=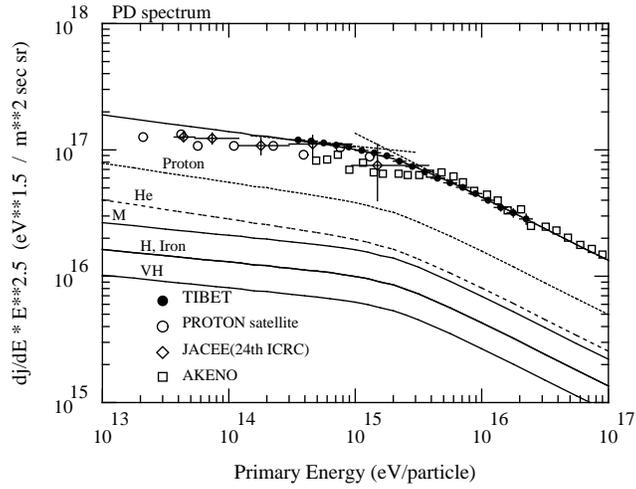,height=11cm} \par
\end{center}

\caption{ Primary cosmic ray composition for  the HD model (top) 
and  the PD model (bottom).  All particle spectrum which is a sum of
each component is normalized to the Tibet data.}
\label{Fig5}
\end{figure}

\begin{figure}
\begin{center}
\vspace{2cm}
  \epsfig{file=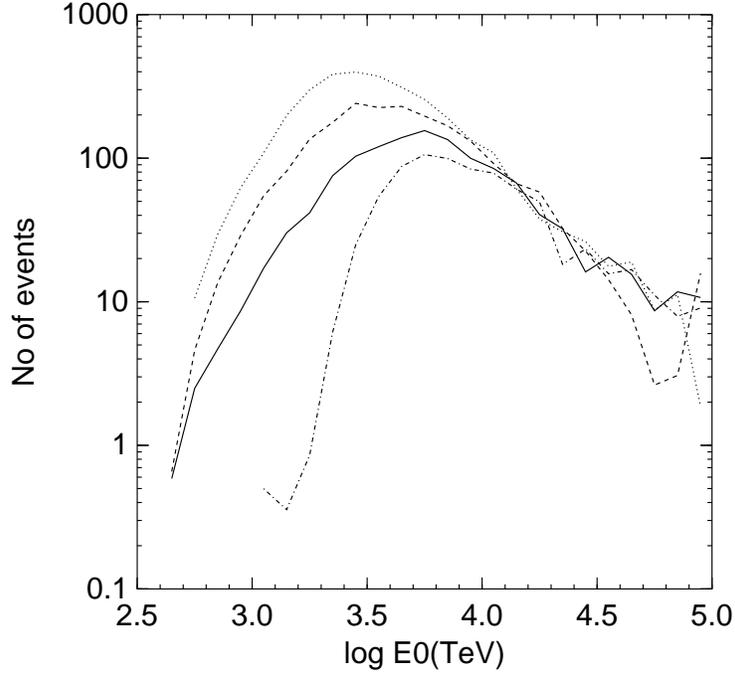,height=9cm} \par
\end{center}

\caption{  Primary spectra of Monte Carlo events selected by  criteria-A 
(see text) for different composition assumptions: pure protons (dotted line), 
PD (dashed line), HD (solid line), and pure irons (dot-dashed line). Monte
Carlo events are generated using CORSIKA-QGSJET.}
\label{Fig6}
\end{figure}

We carried out a hybrid experiment, consisting of the burst 
detector and emulsion chamber
array and the Tibet-II air shower array, at Yangbajing (4300 m above sea level) in Tibet
during the period from 1996 through 1999.
 From this experiment, we observed more than 1000 burst events (high-energy air-shower cores) 
accompanying air showers with energies at the knee region.
Using this data set, we studied the cosmic ray composition at the knee energy region
in comparison with extensive Monte Carlo data.
All the behavior of the observed burst events are shown to be compatible with the
heavily enriched primary composition at the knee. This result suggests that
the mean mass number of the primary particles around 10$^{16}$ eV is
close to silicon or medium nuclei when the composition shown in Fig. \ref{Fig5}(top) is assumed.
 If we combine this with the proton spectrum observed with the same 
experiment \cite{23}, the cosmic composition at the knee region
may be in favor of shock acceleration  at supernova remnants, suggesting a
break of the proton spectrum at energies around 100 TeV.    

We are planning to set up a large-scale  burst array,  consisting of 20 $\times$ 20 
scintillation counters of 0.20 m$^2$ each (50 cm $\times$ 40 cm) which are placed 
at a 1$\sim$2 m grid, 
near the center of the Tibet air shower array.
A 5 cm lead plate may be put on the top of each scintillator to detect the burst events
accompanying air showers. By operating  this new burst array for 1 yr, we may observe 
about 2500 proton-induced events and about 700 iron-induced evens whose primary energies
are in the knee region.  A Monte Carlo study shows that
such burst array can provide information about
each component of the primary cosmic rays at the knee 
with sufficient statistics. This experiment will start within a few years. 

\acknowledgments

This work is supported in part by Grants-in-Aid for Scientific
Research and also for International Science Research from the Ministry
of Education, Science, Sports and Culture in Japan and the Committee
of the Natural Science Foundation and the Academy of Sciences in
China. The support of Japan Society for the Promotion of Science 
 (L.K.D., X.W.X., and C.S.Z.) is also acknowledged.

\noindent
\begin{center}
{\bf APPENDIX \\
 PERFORMANCE OF THE BURST DETECTOR USED IN THIS EXPERIMENT}
\end{center}

 Each burst detector contains a plastic scintillator with the size of 
160 cm $\times$ 50 cm $\times$ 2 cm.  
A PIN PD (HPK S2744-03) with an effective area of 2 cm $\times$ 1 cm was equipped 
at each four corner of each scintillator, as shown in Fig. \ref{FigA1}.
To detect signals from a PD for burst
particles ranging from 10$^3$ to 10$^7$, a preamplifier with an amplification factor of 
260 operating in the frequency range from 17 kHz to 44 MHz (current-current type) was developed. An ADC value from each PD, depending on the size and 
the hit position of a burst (shower) fallen in the
burst detector, can be expressed as  $KN_b(r)$, 
where $r$ is the distance between a PD and the burst position in the 
scintillator, $N_b$ is the burst size
 and $K$  is a constant. Using the ADC values from four corners, we can estimate 
the size and hit position for each burst event using a least-square method.   In this formula, $f(r)$ denotes the 
attenuation of photons in the scintillator.  
In general $f(r)$ can be expressed as exp(-$r/\lambda$) except at small
 distance $r$ and $\lambda$ takes a value around 350 cm for the present scintillator.
Since the size of the burst detector is smaller than the attenuation length,
 errors of the burst hit position  become very large.  So first we slightly polished
 one face  of each scintillator with rough 
sandpaper (No.60) to make  photons scatter randomly 
on this face. Then we found  that $f(r)$ can be well approximated 
 as  ${\large r}^{-\alpha}$ and $\alpha$ $\sim$ 1.1 - 1.2. This relation was confirmed 
by using a nitrogen gas laser and also cosmic ray muons.  
This dependence on the distance $r$ is sufficient to estimate  the burst position in the detector.

We also installed  a calibration unit  which consists of four blue light-emitting diodes (LEDs) each having a 
peak wave length of  450 nm. The LED unit is put on the center of each scintillator
 and is illuminated to transmit light
 through the scintillator to each PD at the corner uniformly, and then all the ADC's
 are calibrated at every   10 min for actual run.  This calibration system 
provides information about    a relative change of ADC values, which may cause 
a large  error for the estimation of  burst  hit positions and burst sizes.

We examined the performance of the burst detector using electron beams of 1.0
GeV/c from the KEK-Tanashi Electron Synchrotron. The beam consisted of 
spills containing about 10$^8$ particles with a time spread of 
about 10 $\mu sec$.
In order to generate  mimic burst events from these bunched beams, 
we randomly extracted part of the particles from the beam spills by adjusting a gate 
width of ADC to 1.2 $\mu s$.  The number of electrons passing through a given gate 
width was estimated by the signals from a probe 
scintillator of 10 cm $\times$ 10 cm placed 
 upstream. Consequently, the electron beams, ranging from several $\times 10^4$ to
$\sim 3 \times 10^5$ per pulse, were vertically exposed to 23 positions on the surface of 
the burst detector as shown in Fig. \ref{FigA2}.

Figure \ref{FigA3} shows the dependence of the ADC values on the distance $r$, obtained
with the electron beams,
where $r$ is the distance between the beam hit position and  PD. The result can be well 
fitted by a power law of $r$, where the number of incident electrons
 measured by the probe scintillator was normalized to 10$^5$ particles.

Using the ADC values from four PDs, the beam positions exposed on the face
of the detector and its intensities (number of electrons) were estimated to
compare with the true ones.
The distribution of the difference between estimated and  actual hit
positions is shown in  Fig. \ref{FigA4}.  We present scatter plots of
 the estimated  and  irradiated number of electrons  in Fig. \ref{FigA5},
 and  the  distribution of the ratio between them is  shown in Fig. \ref{FigA6}.
 From these figures, it is concluded that the hit position of 
a burst in each detector can be estimated with an inaccuracy less 
than 10 cm and
 errors for the size estimation are smaller than  10 \% for the bursts with 
size $ >10^5$ particles.

\newpage 

\begin{figure}
\begin{center}
  \epsfig{file=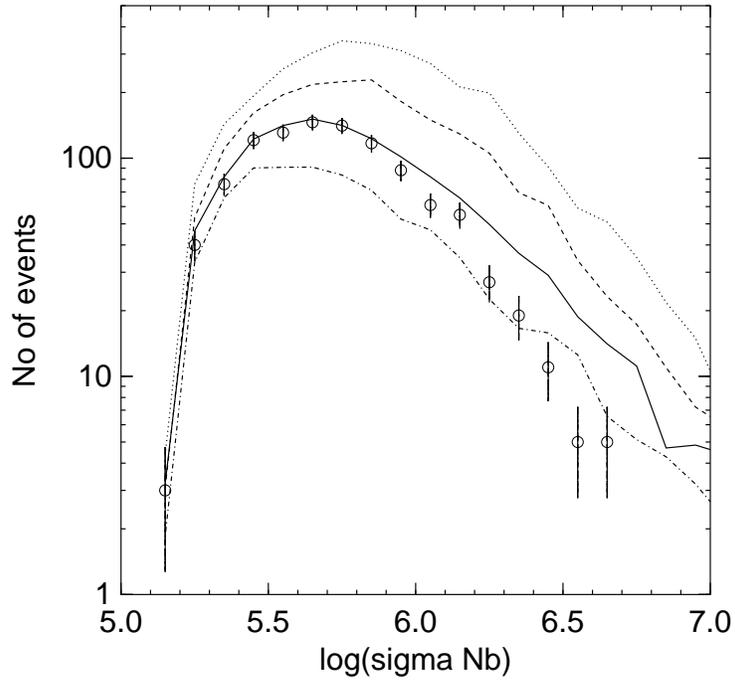,height=9cm} \par
\end{center}

\caption{ Distribution of the total burst size $\sum N_b$. The experimental data
are compared with those of four composition models. Denotations of the 
curves are the same as in Fig. 6. }
\label{Fig7}
\end{figure}

\begin{figure}
\begin{center}
\vspace{2cm}
  \epsfig{file=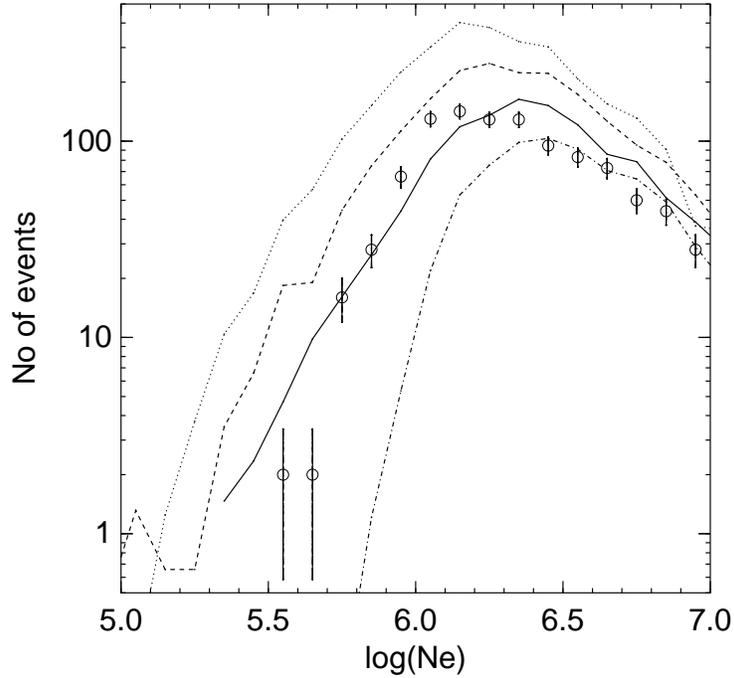,height=9cm} \par
\end{center}

\caption{ Distribution of the air shower size  $N_e$ accompanying the burst event
selected by criteria-A. The experimental data are compared 
with those by different composition models. Denotations of the 
curves are the same as in Fig. 6.}
\label{Fig8}
\end{figure}

\begin{figure}
\begin{center}
  \epsfig{file=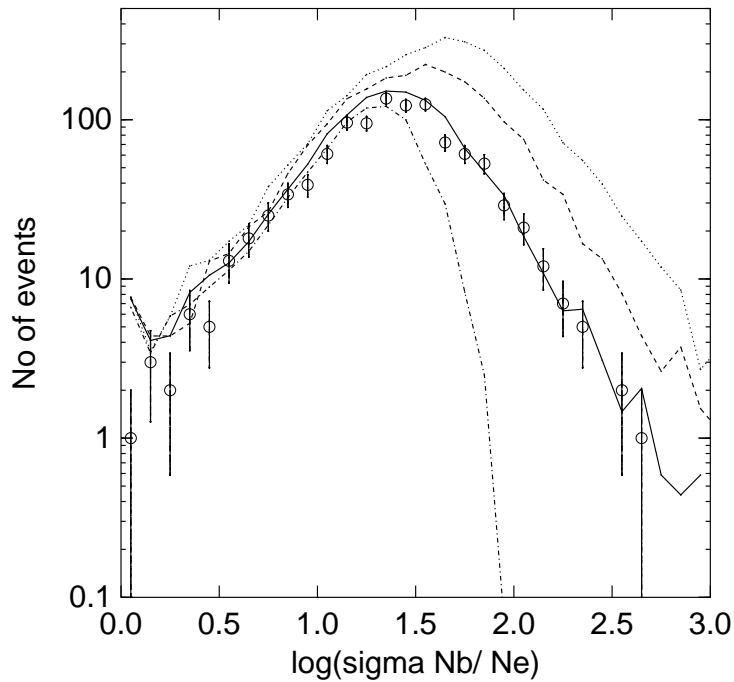,height=9cm} \par
\end{center}

\caption{ Distribution of $(\sum N_b)/N_e$ in comparison with the
Monte Carlo results. Denotations of the curves are the same as in Fig. 6.}
\label{Fig9}
\end{figure}

\begin{figure}
\begin{center}
  \epsfig{file=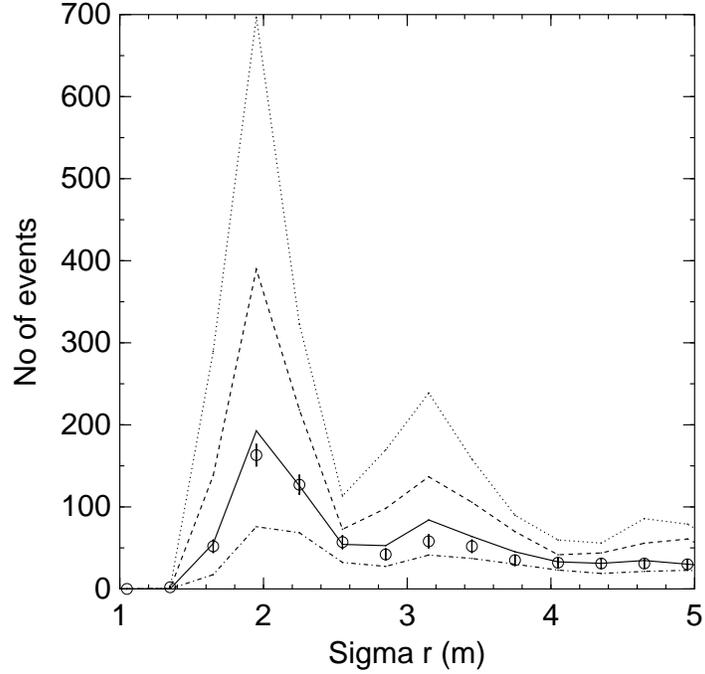,height=9cm} \par
\end{center}

\caption{ Distribution of $\sum r$ for each burst event, where $r$ is a 
distance between the TOP detector position and any one of the other fired 
detectors (see text).
Denotations of the curves are the same as in Fig. 6.}
\label{Fig10}
\end{figure}

\begin{figure}
\begin{center}
  \epsfig{file=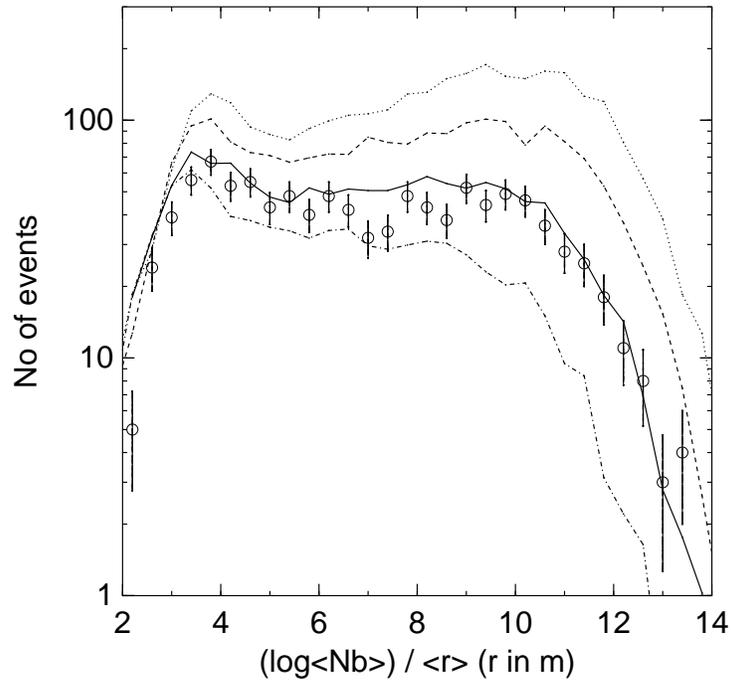,height=9cm} \par
\end{center}

\caption{ Distributions of the ratio  $log \langle N_b \rangle / \langle r \rangle$. The data are compared with the Monte Carlo results.
Denotations of the curves are the same as in Fig. 6.}
\label{Fig11}
\end{figure}

\begin{figure}
\begin{center}
  \epsfig{file=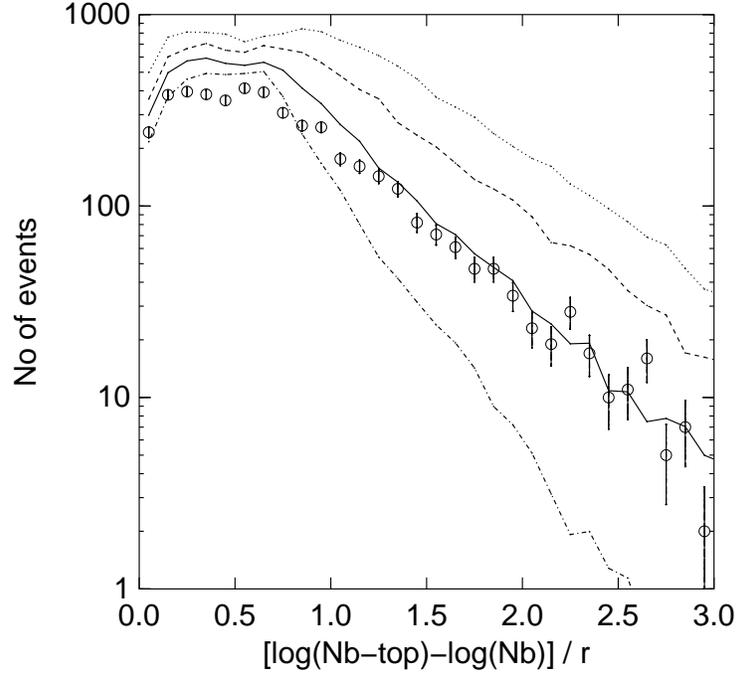,height=9cm} \par
\end{center}

\caption{ Distributions of the burst  spread  expressed
by the gradient parameter $[log(N_b^{top})-log(N_b)] / r$ for each
burst of all events. The data are compared with the Monte Carlo results.
Denotations of the curves are the same as in Fig. 6.}
\label{Fig12}
\end{figure}

\begin{figure}
\begin{center}
  \epsfig{file=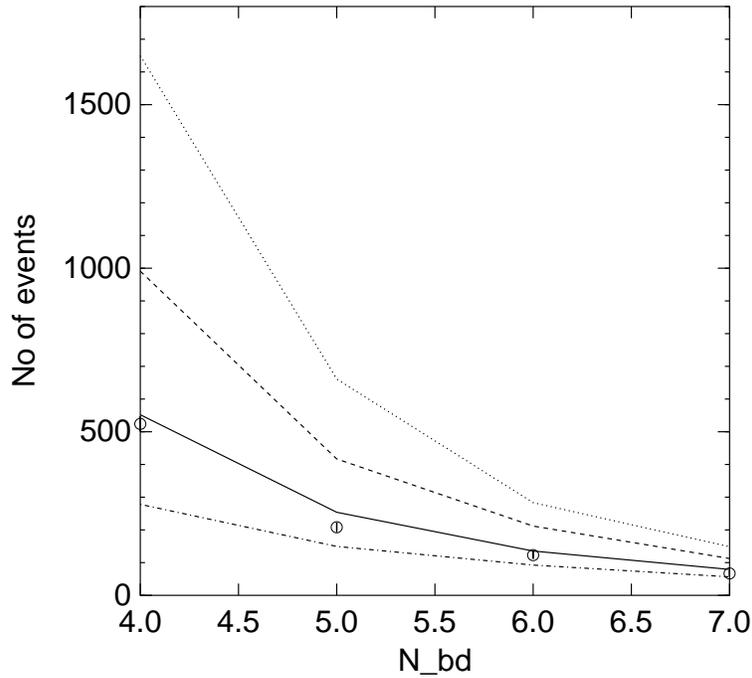,height=9cm} \par
\end{center}

\caption{ Distributions of the number of fired burst detectors $N_{bd}$.
Experimental data  are compared with the Monte Carlo results.
Denotations of curves are the same as in Fig. 6.}
\label{Fig13}
\end{figure}

\begin{figure}
\begin{center}
  \epsfig{file=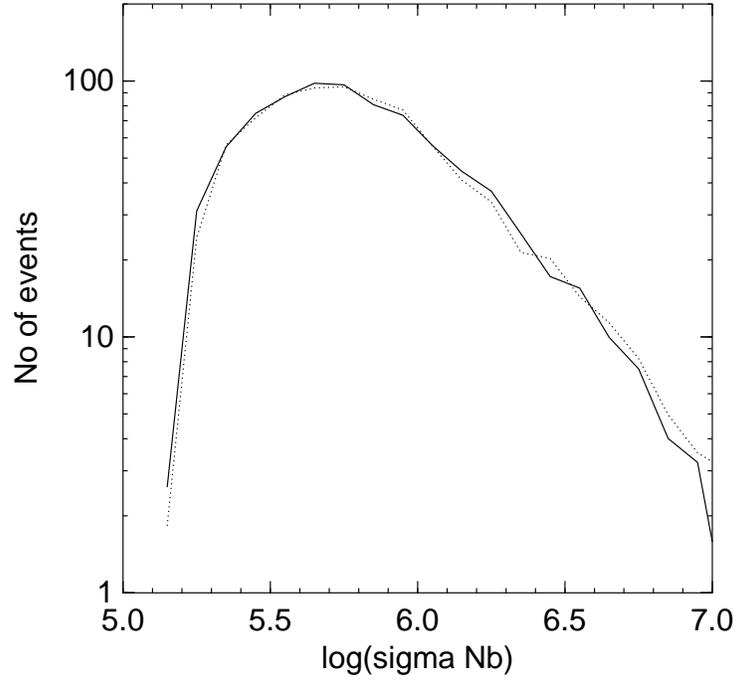,height=9cm} \par
\end{center}

\caption{ Comparison of distributions of the total burst size $\sum N_b$,
obtained  by CORSIKA-QGSJET (solid line) and COSMOS (dotted line).}
\label{Fig14}
\end{figure}

\begin{figure}
\begin{center}
  \epsfig{file=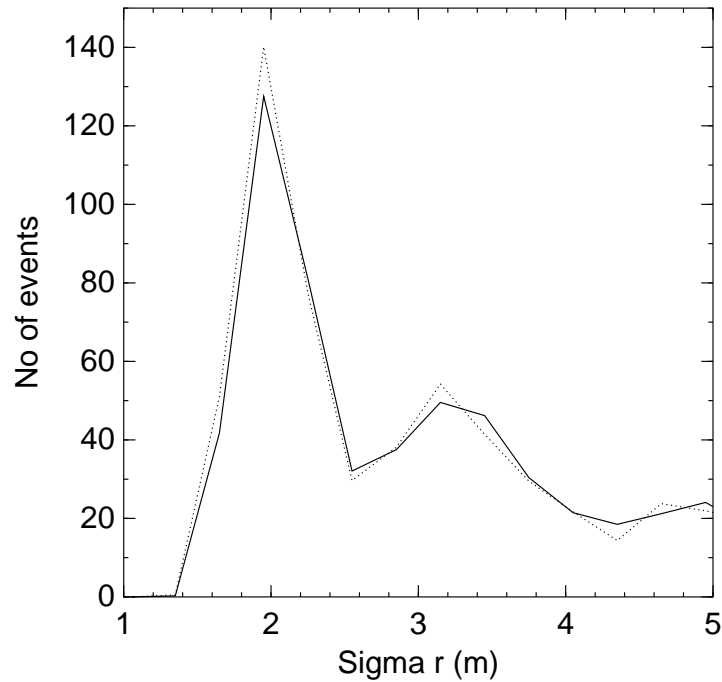,height=9cm} \par
\end{center}

\caption{ Comparison of distributions of the sum of distance between bursts for each event $\sum r$, obtained  by CORSIKA-QGSJET (solid line)
and COSMOS (dotted line).}
\label{Fig15}
\end{figure}

\begin{figure}
\begin{center}
  \epsfig{file=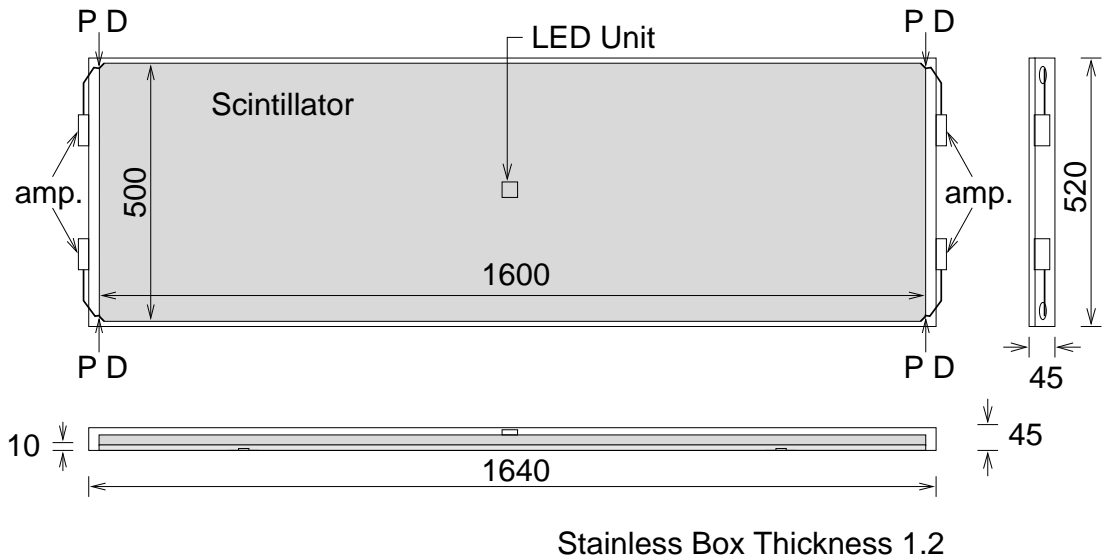,height=8cm} \par
\end{center}

\caption{ Schematic view of the burst detector used in this experiment. Numerals shown in the figure are in units of mm.}
\label{FigA1}
\end{figure}

\begin{figure}
\begin{center}
  \epsfig{file=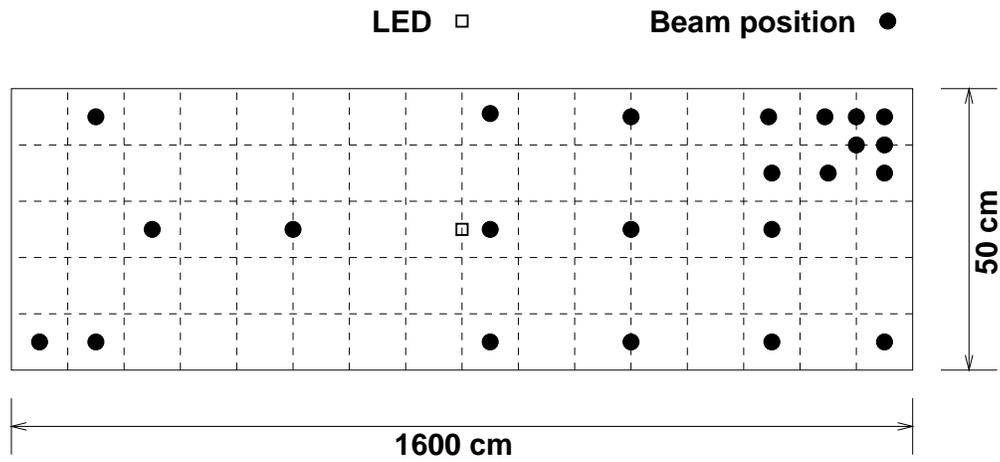,height=6cm} \par
\end{center}

\caption{ Beam hit positions on the surface of the detector.}
\label{FigA2}
\end{figure}

\begin{figure}
\begin{center}
  \epsfig{file=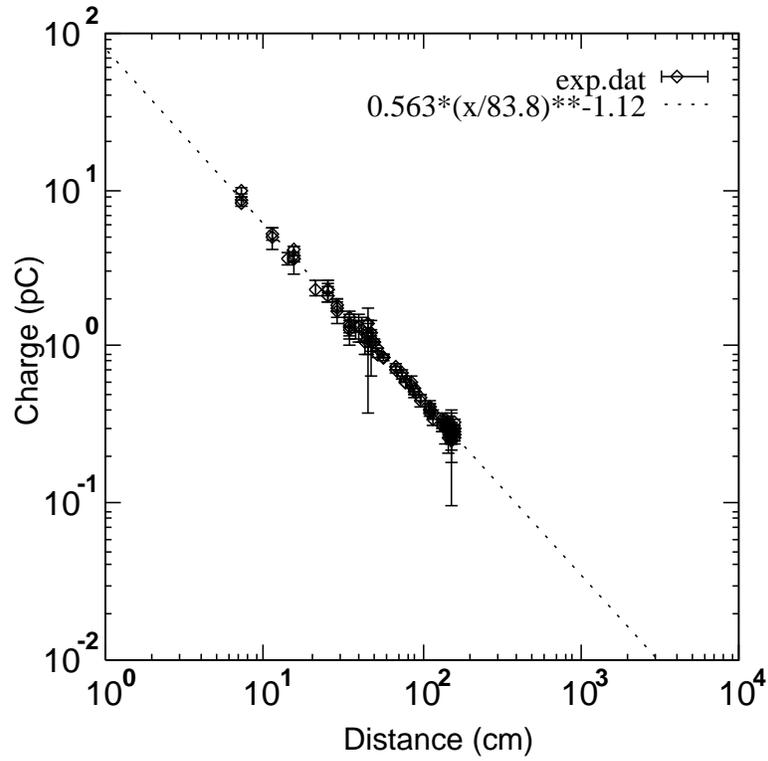,height=10cm} \par
\end{center}

\caption{ Attenuation of photons in the scintillator used for the burst detector, obtained
using electron beams.}
\label{FigA3}
\end{figure}

\begin{figure}
\begin{center}
  \epsfig{file=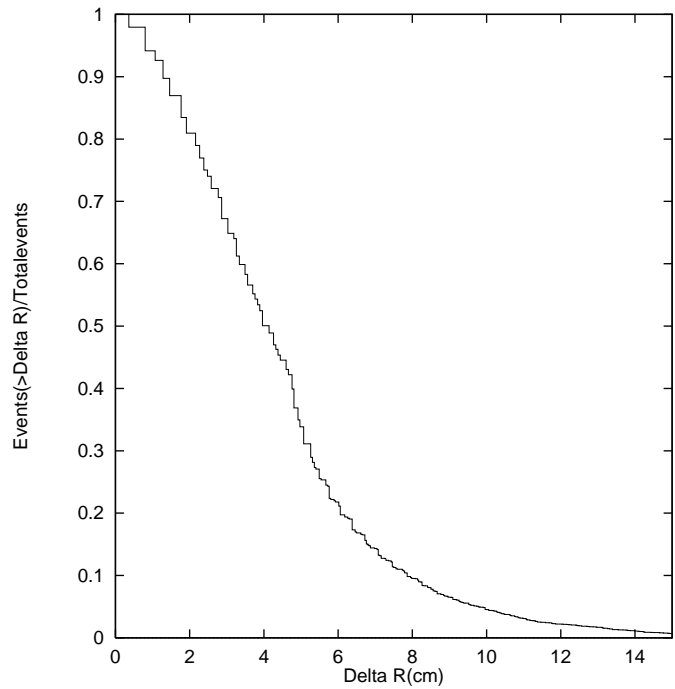,height=9cm} \par
\end{center}

\caption{ Distribution (integral) of the difference  between  estimated and irradiated
positions.}
\label{FigA4}
\end{figure}

\begin{figure}
\begin{center}
  \epsfig{file=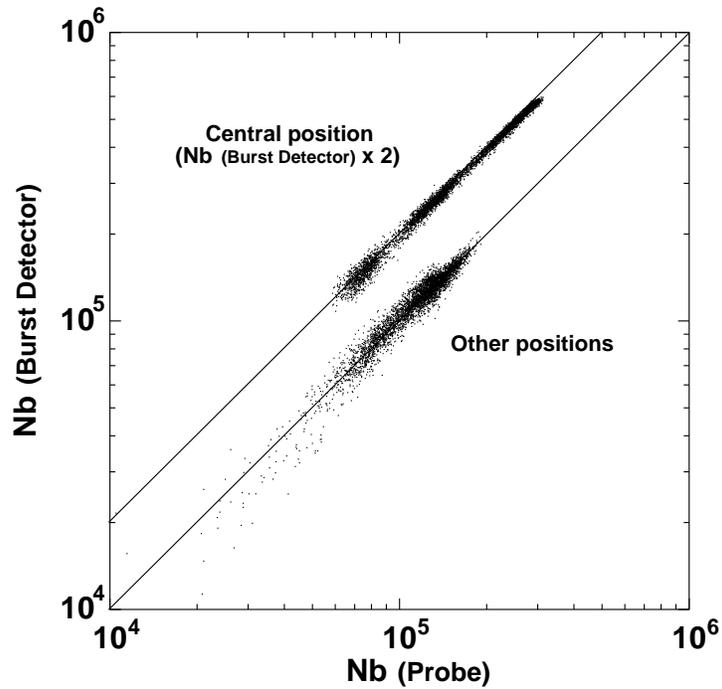,height=9cm} \par
\end{center}

\caption{ Scatter plots of  estimated  and  irradiated number of electrons. 
The number of electrons at various beam positions on the face of
the detector is normalized to $10^5$ electrons.}
\label{FigA5}
\end{figure}

\begin{figure} 
\begin{center}
  \epsfig{file=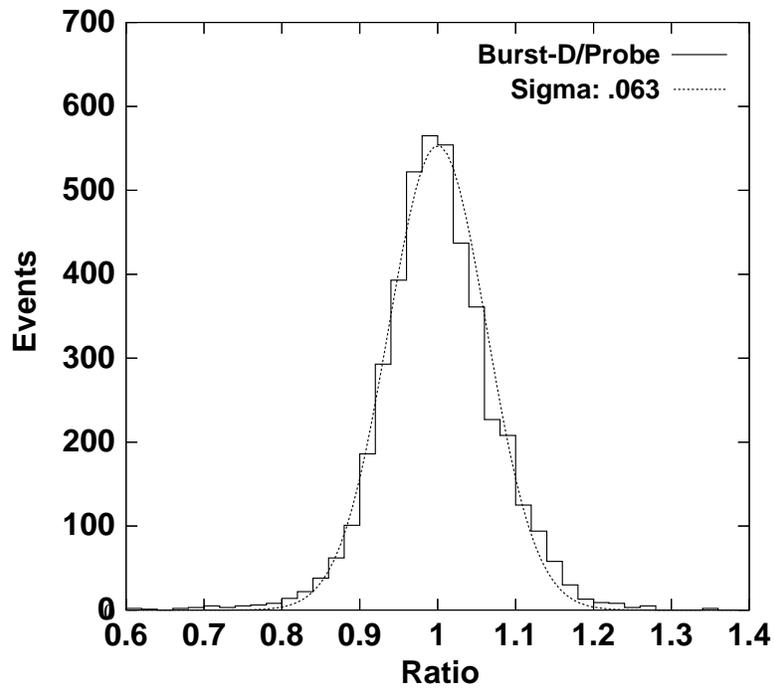,height=9cm} \par
\end{center}

\caption{ Distribution of the ratio of estimated  and irradiated number of electrons
shown in Fig. \ref{FigA5}. Dotted line is a Gaussian fit.}
\label{FigA6}
\end{figure}

\end{document}